\documentstyle[11pt]{article}
\newcommand{\vsig}{\mbox{\boldmath $\sigma$ \unboldmath}}
\newcommand{\veps}{\mbox{\boldmath $\epsilon$ \unboldmath}}
\newcommand{\valf}{\mbox{\boldmath $\alpha$ \unboldmath}}

\begin{document}
\bibliographystyle{unsrt}

\title{\bf Vector Meson Photoproduction with an Effective Lagrangian
 in the Quark Model I: Formalism}

\author{Qiang Zhao, Zhenping Li \\  
Department of Physics, Peking University, Beijing, 100871, P.R.China\\
C. Bennhold\\
Department of Physics, Center for Nuclear Studies,\\
 The George Washington University, Washington, D.C., 20052, USA}  
\maketitle  
  
\begin{abstract}
A quark model approach to the photoproduction of vector mesons off nucleons is  
proposed.  Its starting point is an effective Lagrangian of the interaction  
 between the vector meson and the quarks inside the baryon.  This paper 
develops the formalism and  highlights the
dynamical roles of s-channel resonances.
It is demonstrated that the proposed approach is ideal to support
searching for the missing resonances in the vector
meson photoproduction experiments proposed  at TJNAF.
\end{abstract}

\section*{\bf 1. Introduction}
 The newly established electron and photon facilities have made
it possible to investigate the photoproduction of
vector mesons, such as $\omega$ and $\rho$,
on the nucleon with much better energy and angular
resolutions.  Such experiments have already been carried out at ELSA\cite{elsa},
and will be
performed at TJNAF\cite{cebaf} in the near future. The data from these measurements
should be able to help us resolve the puzzle of the so-called
``missing resonance``\cite{NRQM,capstick}.  A better understanding of the
reaction mechanism is therefore crucial
in order to extract resonance information from  vector meson photoproduction
observables. Historically, one of the attempts to understand
$\rho$ photoproduction  in the reaction 
$\gamma p\to \pi^+\pi^- p$ has lead to the so-called ``S\"oding 
Model``\cite{soding}, which explains low-energy $\rho^0$ production 
mainly through t-channel exchange. Although the S\"oding Model has had some
 success in reproducing the data in certain kinematic regions
it cannot be used to study the dynamical role of the s-channel 
resonances in vector meson photoproduction. A more systematic approach 
that includes the t-channel exchange terms along with the s- and u-channel
resonances is needed to meet the challenge of new high precision data
and serve the purpose of searching for
``missing resonances``. 

The constituent quark model approach has been shown to very successfully reproduce data in the area of pseudoscalar meson photoproduction~\cite{pseudoscalar} with a minimal 
number of free parameters since it introduces the quark degrees of 
freedom directly into the reaction mechanism.
More importantly, it allows to study
the internal structure of the contributing s-channel resonances
directly in connection with the reaction mechanism.
It is therefore natural to extend the quark model 
approach to vector meson photoproduction in the resonance region, 
the development of the formalism for that will be the focus of this paper.
  
A major difference for vector meson production from the pseudoscalar meson case in the quark model is that the interaction between the vector mesons and the quarks inside the baryon
 is largely unknown.  Although phenomenological models have been developed
to evaluate baryon resonance 
decays into a nucleon and a vector meson,  
 such as the quark pair creaction model~\cite{yaouanc} or the $^3P_0$ model,
these approaches are unsuitable for the description of 
vector meson photoproductions. This is due to the fact that they
 only yield transition amplitudes for s-channel resonances, but
contain no information on how to derive the non-resonant terms in the u- and 
t-channels.  Therefore,  we choose an effective Lagrangian 
here as our starting point that satisfies the fundamental 
symmetries and determines the transition amplitudes not only in the s-channel but 
also in the t- and u- channels.

Even though the effective Lagrangians are different for pseudoscalar and 
vector meson photoproductions, the implementation follows the same 
guidelines.
The transition amplitudes for each resonance in the s-channel
below 2 GeV will be included
 explicitly, while the resonances above 2 GeV for a given quantum number $n$ 
in the harmonic oscillator basis of the quark model
are treated as degenerate, so that their transition amplitudes can be written 
in a compact form. Similarly, the excited resonances in the u-channel are 
assumed to be degenerate as well. Only the mass splitting between the 
spin 1/2 and spin 3/2 resonances with $n=0$ in the harmonic oscillator basis,
such as the splitting between nucleon and $\Delta$ resonance, is
found to be significant, thus, 
the transition amplitudes for the spin 3/2 resonance with $n=0$ in 
the u-channel will be included separately.

The effective Lagrangian employed here generates not only the s- and u-channel
exchanges but also a t-channel term containing the vector meson.
For charged vector mesons gauge invariance also mandates a seagull term.
An open question in this approach is whether additional
 t-channel contribution, such as the Pomeron exchange, responsible for the diffractive behavior in the small $t$ region is needed. Previous studies in the pseudoscalar sector suggest that additional t-channel 
exchanges were not required to describe the data.
This could be understood via application of the duality hypothesis~\cite{duality}. 
However, whether it is applicable for vector meson photoproduction is not clear at present,
this will be investigated further by comparing the numerical results 
of the model with experimental data. 

In Section 2, we briefly discuss some of the observables
used in our approach, which have been developed extensively in Ref. 
\cite{tabakin}.  The effective Lagrangian for the quark-meson interaction
is presented in Section 3, along with detail expressions for the the s-, u- 
and t-channel contributions arising from this Lagrangian.  
Finally, conclusions will be presented in the Section 4.
  
\section*{\bf 2. Observables and Helicity Amplitudes }  
  
Before presenting our quark model approach we introduce some general features
of vector meson photoproduction on the nucleon.
The basic amplitude $\cal F$ for $\gamma + N \to V+N$ is defined as
\begin{equation}  
{\cal F}=\langle {\bf q}\lambda_V\lambda_2|T|{\bf k}
 \lambda\lambda_1\rangle,  
\label{1}  
\end{equation}  
where ${\bf k}$ and ${\bf q}$ are the momenta of the incoming 
photon and outgoing vector meson.
The helicity states are denoted by $\lambda=\pm 1$ 
for the incident photon,   
$\lambda_V=0,\pm 1$ for the outgoing  vector meson, 
and $\lambda_1=\pm 1/2$,$\lambda_2=\pm 1/2$  
for the initial and final state nucleons, respectively. Following 
Ref.~\cite{tabakin}, the amplitude $\cal F$
can be expressed as a $6\times 4$ matrix in the helicity space,
\begin{equation}  
{\cal F}=\left( \begin{array}{cccc}  
H_{21} & H_{11} & H_{3-1} & -H_{4-1}\\  
H_{41} & H_{31} & -H_{1-1} & H_{2-1}\\
H_{20} & H_{10} & -H_{30} & H_{40}\\
H_{40} & H_{30} & H_{10} & -H_{20}\\
H_{2-1} & H_{1-1} & H_{31} & -H_{41}\\
H_{4-1} & H_{3-1} & -H_{11} & H_{21}\\ 
\end{array} \right ).
\label{2}
\end{equation}  
Because of the parity conservation,  
\begin{equation}  
\langle {\bf q}\lambda_V\lambda_2|T|{\bf k} \lambda\lambda_1\rangle=  
(-1)^{\Lambda_f-\Lambda_i}\langle {\bf q} -\lambda_V -\lambda_2|T|{\bf k}-  
\lambda -\lambda_1\rangle,  
\end{equation}
where $\Lambda_f=\lambda -\lambda_1$ and $\Lambda_i=\lambda_V -\lambda_2$
 in the Jacob-Wick(JW) convention,  
the $ H_{a\lambda_V}(\theta)$ in Eq.(\ref{2}) reduces to 12 independent  
 complex helicity amplitudes:

\begin{eqnarray} \label{helicity}
H_{1\lambda_V}&=  &\langle \lambda_V, \lambda_2=+1/2|T|\lambda=1,   
\lambda_1=-1/2\rangle\nonumber\\  
H_{2\lambda_V}&=  &\langle \lambda_V, \lambda_2=+1/2|T|\lambda=1,   
\lambda_1=+1/2\rangle\nonumber\\  
H_{3\lambda_V}&=  &\langle \lambda_V, \lambda_2=-1/2|T|\lambda=1,
\lambda_1=-1/2\rangle\nonumber\\  
H_{4\lambda_V}&=  &\langle \lambda_V, \lambda_2=-1/2|T|\lambda=1,  
\lambda_1=+1/2\rangle.  
\end{eqnarray}  

Each experimental observable $\Omega$ can be written in the general 
{\sl bilinear helicity product}(BHP) form,
\begin{eqnarray}
\check{\Omega}^{\alpha\beta}&=&\Omega^{\alpha\beta}{\cal T}(\theta)\nonumber\\
&=&\pm\frac 12\langle H|\Gamma^\alpha\omega^\beta|H \rangle\nonumber\\
&=&\pm\frac 12\sum_{a,b,\lambda_V,\lambda^\prime_V} H^*_{a\lambda_V}\Gamma^\alpha_{ab}\omega^\beta_{\lambda_V\lambda^\prime_V}H_{b
\lambda^\prime_V}.
\end{eqnarray}
For example, the differential cross 
section operator is given by:
\begin{eqnarray}
\check{\Omega}^{\alpha=1,\beta=1}&=&{\cal T}(\theta)\nonumber\\
&=&\frac 12\langle H|\fbox{$\Gamma^1$}\fbox{$\omega^1$}|H\rangle\nonumber\\
&=&\frac 12\sum^4_{a=1}\sum_{\lambda_V=0,\pm 1} |H_{a\lambda_V}|^2,
\end{eqnarray}
where the box
 frames denote the diagonal structure of the matrices. The
$\Gamma$ and $\omega$ matrices labeled by different $\alpha$ and $\beta$ 
correspond to different spin observables. With the phase space factor, the 
differential cross section has the expression,
\begin{eqnarray}
\frac{d\sigma}{d\Omega_{c.m.}}&=&(P.S. factor){\cal T}(\theta)\nonumber\\
&=&\frac{\alpha_e \omega_m(E_f+M_N)(E_i+M_N)}{8\pi s}{|{\bf q}|}\frac 12
\sum^4_{a=1}\sum_{\lambda_V=0,\pm 1} |H_{a\lambda_V}|^2
\end{eqnarray}
in the center of mass frame, where $\sqrt{s}$ is the total energy of the
 system, $\mu$ and $M_N$ represent
the masses of the outgoing meson and nucleon,  
and $\omega$, $\omega_m$ denote the energy of the photon and meson,
 respectively.

These helicity amplitudes are usually related to the density matrix 
elements $\rho_{ik}$~\cite{schilling}, which are measured by the experiments~\cite{Ballam}. 
They are defined as:  
\begin{eqnarray}  
\rho^0_{ik}&=  &\frac{1}{A}\sum_{\lambda\lambda_2\lambda_1}H_{\lambda_{V_i}  
\lambda_2,  
\lambda\lambda_1} H^*_{\lambda_{V_k}\lambda_2, \lambda\lambda_1},  
\nonumber\\
\rho^1_{ik}&=  &\frac{1}{A}\sum_{\lambda\lambda_2\lambda_1}  
H_{\lambda_{V_i}\lambda_2,  
-\lambda\lambda_1} H^*_{\lambda_{V_k}\lambda_2, \lambda\lambda_1},  
\nonumber\\  
\rho^2_{ik}&=  &\frac{i}{A}\sum_{\lambda\lambda_2\lambda_1}\lambda   
H_{\lambda_{V_i}\lambda_2, -\lambda\lambda_1} H^*_{\lambda_{V_k}\lambda_2,   
\lambda\lambda_1},\nonumber\\  
\rho^3_{ik}&=  &\frac{i}{A}\sum_{\lambda\lambda_2\lambda_1}\lambda   
H_{\lambda_{V_i}\lambda_2, \lambda\lambda_1} H^*_{\lambda_{V_k}\lambda_2,   
\lambda\lambda_1},   
\end{eqnarray}  
where   
\begin{equation}  
A=\sum_{\lambda_{V_i}\lambda\lambda_2\lambda_1} H_{\lambda_{V_i}\lambda_2,  
\lambda\lambda_1} H^*_{\lambda_{V_k}\lambda_2, \lambda\lambda_1},  
\end{equation}  
where $\rho_{ik}$ stands for $\rho_{\lambda_{V_i}\lambda_{V_k}}$, and $\lambda_{V_i}$, $\lambda_{V_k}$ denote the helicity of the produced vector mesons.

For example, the angular distribution  
 for $\rho^0$ decaying into $\pi^+\pi^-$ 
produced by linearly polarized photons can be expressed  
 in terms of nine independent measurable spin-density matrix elements
\begin{eqnarray}  
W(cos\theta, \phi,\Phi)&=  &\frac 3{4\pi}[\frac 12(1-\rho^0_{00})  
+\frac 12(3\rho^0_{00}-1)cos^2\theta-\sqrt2Re\rho^0_{10}sin2  
\theta cos\phi\nonumber\\
&&-\rho^0_{1-1}sin^2\theta cos2\phi  
-P_\gamma cos2\Phi(\rho^1_{11}sin^2\theta+\rho^1_{00}cos^2  
\theta\nonumber\\
&&-\sqrt2 Re\rho^1_{10}sin2\theta cos\phi-\rho^1_{1-1}sin^2  
\theta cos2\phi)\nonumber\\
&&-P_\gamma sin2\phi(\sqrt2 Im\rho^2_{10}sin2\theta sin\phi+Im\rho^2_{1-1}sin^2\theta sin2\phi)],  
\end{eqnarray}  
where $P_\gamma$ is the degree of the linear polarization of the  
 photon, $\Phi$ is the angle of the photon electric polarization  
 vector with respect to the production plane measured in the c.m.  
 system, and $\theta$ and $\phi$ are the polar and azimuthal angles   
of the $\pi^+$ which is produced by the $\rho^0$ decay in the   
$\rho^0$ rest frame.

\section*{\bf 3. Quark Model Approach for Vector Meson Photoproduction}  
  
The starting point of the quark model approach is the effective Lagrangian,
\begin{equation} \label{3.0}  
L_{eff}=-\overline{\psi}\gamma_\mu p^\mu\psi+\overline{\psi}\gamma_
\mu e_qA^\mu\psi +\overline{\psi}(a\gamma_\mu +  
\frac{ib\sigma_{\mu\nu}q^\nu}{2m_q}) \phi^\mu_m \psi,
\end{equation}  
where the quark field $\psi$ is expressed as  
\begin{equation}  
\psi =\left( \begin{array}{c}  
\psi (u)\\ \psi (d) \\ \psi (s) \end{array} \right ),   
\end{equation}  
and the meson field $\phi^\mu_m $ is a 3$\otimes$3 matrix,  
\begin{equation}  
\phi_m =\left( \begin{array}{ccc}  
\frac{1}{\sqrt{2}}\rho^{0}+\frac{1}{\sqrt{2}}\omega & \rho^{+} & K^{*+}\\  
\rho^{-} & -\frac{1}{\sqrt{2}}\rho^{0}+\frac{1}{\sqrt{2}}\omega & K^{*0}\\  
K^{*-} & \overline{K}^{*0}  &\phi  
\end{array} \right )   
\end{equation}  
in which the vector mesons are treated as point-like particles. 
At tree level, the transition matrix element based on the effective   
Lagrangian in Eq.(\ref{3.0}) can be written as the sum of contributions  
 from the s-, u- and t- channels,
\begin{equation}  
M_{fi}=M^s_{fi}+M^u_{fi}+M^t_{fi},  
\label{3.1}  
\end{equation}  
where the s- and u-channel contributions in Eq.(\ref{3.1}) 
 have the following form,  
\begin{eqnarray}  
M^s_{fi}+M^u_{fi}&=  &\sum_{j} \langle N_f|H_m|N_j\rangle\langle   
N_j|\frac{1}{E_i+\omega-E_j}H_e|N_i\rangle\nonumber\\  
&&+\sum_{j} \langle N_f|H_e\frac{1}{E_i-\omega_m-E_j}
|N_j\rangle\langle N_j|H_m|N_i\rangle,  
\end{eqnarray}  
where the electromagnetic coupling vertex is  
\begin{equation}  
H_e=-\overline{\psi}\gamma_\mu e_qA^\mu\psi,  
\end{equation}  
and the quark-meson coupling vertex is  
\begin{equation} \label{Hm} 
H_m=-\overline{\psi}(a\gamma_\mu +\frac{ib\sigma_{\mu\nu}  
q^\nu}{2m_q}) \phi^\mu_m \psi,  
\end{equation}  
where $m_q$ in is the quark mass and the constants $a$ and $b$ in Eq.(\ref{3.0})
and (\ref{Hm}) are the vector and tensor   
coupling constants, which will be treated as free 
parameters in our approach. The initial and final
states of the nucleon are denoted by $|N_i\rangle$ and $|N_f\rangle$, 
respectively, and $|N_j\rangle$
 is the intermediate resonance state while $E_i$ and $E_j$ 
are the energies of the inital nucleon and the intermediate resonance. 

An important test of the transition matrix elements 
\begin{equation}  
M_{fi}=\langle \lambda_2|J_{\mu\nu}\epsilon^\mu
\epsilon^\nu_m|\lambda_1\rangle,  
\label{2.1}  
\end{equation}  
would be gauge invariance:  
\begin{equation}  
\langle \lambda_2|J_{\mu\nu}k^\mu|\lambda_1\rangle=\langle  
 \lambda_2|J_{\mu\nu} q^\nu_m|\lambda_1\rangle=0,  
\label{2.2}  
\end{equation}  
where $\epsilon^\mu_m$, $\epsilon^\nu$ are the polarization  
 vectors of the vector mesons and photons.
However, we find that the condition $\langle  
 \lambda_2|J_{\mu\nu} q^\nu_m|\lambda_1\rangle=0$ is not satisfied for
the t-channel vector meson exchange term,
\begin{equation}  
M^t_{fi}=a\langle N_f|\sum_{l}\frac{e_m}{2q\cdot k}  
\{2q\cdot\epsilon\gamma\cdot\epsilon_m-\gamma\cdot   
q\epsilon\cdot\epsilon_m+k\cdot\epsilon_m\gamma\cdot\epsilon\}  
e^{i({\bf k}-{\bf q})\cdot{\bf r}_l}| N_i\rangle,  
\label{3.3}  
\end{equation} 
based on the Feynman rules for the photon-vector meson coupling, and
the relation
\begin{equation}  
\langle N_f|\gamma\cdot(q-k) e^{i({\bf k}-{\bf q})  
\cdot{\bf r}_l}| N_i\rangle=0.  
\end{equation}  
To remedy this problem, we add a gauge fixing term, so that 
\begin{equation}  
M^t_{fi}=a\langle N_f|\sum_{l}\frac{e_m}{2q\cdot k}  
2\{q\cdot\epsilon\gamma\cdot\epsilon_m-\gamma\cdot q\epsilon\cdot\epsilon_m+k\cdot\epsilon_m\gamma\cdot\epsilon\}  
e^{i({\bf k}-{\bf q})\cdot{\bf r}_l}| N_i\rangle.  
\end{equation}

The techniques of deriving the transition amplitudes have been developed
for Compton scattering~\cite{Compton}. Follow the same procedure as given in Eq.(14) of Ref.~\cite{pseudoscalar}, we can divide the photon interaction into two parts and the contributions 
from the s- and u- channel can be rewritten as:
\begin{eqnarray}  
M^{s+u}_{fi}&=&i\langle N_f|[g_e,H_m]|N_i\rangle \nonumber\\  
&&+i\omega\sum_{j}\langle N_f|H_m|N_j\rangle\langle   
N_j|\frac{1}{E_i+\omega-E_j}h_e|N_i\rangle\nonumber\\  
&&+i\omega\sum_{j} \langle N_f|h_e\frac{1}{E_i-\omega_m-E_j}  
|N_j\rangle\langle N_j|H_m|N_i\rangle, 
\label{3.2}  
\end{eqnarray}  
where  
\begin{equation}  
g_e=\sum_{l}e_l{{\bf r}_l\cdot{\bf\epsilon}}e^{i{\bf k\cdot r}_l},  
\end{equation}  
\begin{equation}  
h_e=\sum_{l}e_l{{\bf r}_l\cdot{\bf\epsilon}}(1-\valf\cdot  
{\hat{\bf k}})e^{i{\bf k\cdot r}_l}  
\end{equation}  
\begin{equation}
{\bf{\hat k}}=\frac{\bf k}{\omega}.
\end{equation}
 
The first term in Eq.(\ref{3.2}) can be identified as a seagull term;
it is proportional to the charge of the outgoing vector meson.
The second and third term in Eq.(\ref{3.2}) represents the s- and u-channel
contributions. Adopting the same strategy as in the pseudoscalar case,
 we include a complete set of helicity amplitudes for each
of the s-channel resonances below 2GeV in the $SU(6)\otimes O(3)$
 symmetry limit. The resonances above 2GeV
are treated as degenerate in order to express each contribution from  
all resonances with quantum number $n$ in a compact form.
The contributions from the resonances with the largest spin for  
 a given quantum number $n$ were found to be  the most important as the energy   
increases~\cite{pseudoscalar}.   
This corresponds to spin $J=n+1/2$ with $I=1/2$ for the reactions  
 $\gamma N\to K^*\Lambda$ and $\gamma N\to \omega N$,  
 and $J=n+3/2$ with $I=3/2$ for the reactions $\gamma N\to K^*\Sigma$ and  
 $\gamma N\to \rho N$.

Similar to the pseudoscalar case, the contributions from   
the u-channel resonances are divided into two parts as well. The
first part contains the resonances with   
the quantum number $n=0$, which include the spin 1/2 states,  
 such as the $\Lambda$, $\Sigma$ and the nucleons, 
and the spin 3/2 resonances, such as the $\Sigma^*$ in $K^*$ photoproduction
and $\Delta(1232)$ resonance in $\rho$ photoproduction. Because   
the mass splitting between the spin 1/2 and spin 3/2 resonances   
for $n=0$ is significant, they have to be treated separately.   
The transition amplitudes for these u-channel resonances will  
also be written in terms of the helicity amplitudes.
 The second part comes from the excited   
resonances with quantum number $n\ge 1$. As the contributions  
 from the u-channel resonances are not sensitive to the precise 
mass positions, they can be treated as degenerate as well,
so that the contributions from these resonances can again
be written in a compact form.  

\subsection*{\bf 3.1. The seagull term }  

The transition amplitude is divided into the transverse
and longitudinal amplitudes according to the polarization of the outgoing 
vector mesons. The longitudinal polarization vector for a vector meson with
mass $\mu$ and momentum ${\bf q}$ is,
\begin{equation}    
\epsilon^\mu_L =\frac{1}{\mu} \left( \begin{array}{c}    
|{\bf q}|\\ \omega_m \frac{\bf q}{|{\bf q}|} \end{array} \right )     
\end{equation}    
where $\omega_m=\sqrt{{\bf q}^2+\mu^2}$ is the energy of the outgoing 
vector mesons.
Thus, the longitudinal interaction at the quark-meson vertex 
can be written as    
\begin{equation}    
H^L_m=\epsilon^\mu_L J_\mu=\epsilon_0J_0-\epsilon_3J_3    
\end{equation}    
where $\epsilon_3$ corresponds to the direction of the momentum ${\bf q}$.
The transition amplitudes of the s- and u-channel for the 
longitudinal quark-meson coupling become,    
\begin{eqnarray}    
M^{s+u}_{fi}(L)&=& i\langle N_f|[g_e,H^L_m]|N_i\rangle \nonumber \\    
&&- i\omega \langle N_f|[h_e,\frac{\epsilon_3}{q_3}J_0]|N_i\rangle  
\nonumber \\    
&&+i\omega\sum_{j}\langle N_f|(\epsilon_0-\frac{\omega_m}{q_3}\epsilon_3)J_0|N_j\rangle
\langle N_j|\frac{1}{E_i+\omega-E_j}h_e|N_i\rangle\nonumber \\    
&&+i\omega\sum_{j} \langle N_f|h_e\frac{1}{E_i-\omega_m-E_j}|N_j  
\rangle\langle N_j|(\epsilon_0-\frac{\omega_m}{q_3}\epsilon_3)J_0|N_i
\rangle,    
\label{4.2})   
\end{eqnarray}    
where the first two terms are seagull terms which 
are cancelled by similar terms in the t-channel.
 The corresponding expressions for the t-channel amplitudes
are given in the appendix; the last two terms will be discussed in the 
following sections.

The nonrelativistic expansion of the transverse meson quark
interaction vertex gives,    
\begin{equation}    
H^T_m=\sum_{l} \{i\frac{b^\prime}{2m_q}\vsig_l\cdot({\bf q}
\times\veps_v)  
+a{\bf A}\cdot\veps_v+\frac{a}{2\mu_q}{\bf p}^\prime_l  
\cdot\veps_v\}{\hat I}_le^{-i{\bf q\cdot r}_l}     
\end{equation}    
where $b^\prime=b-a$ and 
\begin{equation}  
{\hat I}_l =\left\{ \begin{array}{ccc}
a^\dagger _l(s)a_l(u) & \qquad\mbox{for}\qquad & K^{*+}\\ 
a^\dagger_l(s)a_l(d) & \qquad\mbox{for}\qquad & K^{*0}\\
a^\dagger_l(d)a_l(u) & \qquad\mbox{for}\qquad & \rho^+\\
-\frac{1}{\sqrt 2}(a^\dagger _l(u)a_l(u) -a^\dagger _l(d)a_l(d)) & \qquad\mbox{for}\qquad & \rho^0\\
1 &\qquad\mbox{for}\qquad & \omega 
\end{array} \right.   
\end{equation}  
The vector ${\bf A}$ has the general form, 
\begin{equation}    
{\bf A}=\frac{{\bf P}_f}{E_f+M_f}+\frac{{\bf P}_i}{E_i+M_i},    
\end{equation}    
which comes from the center-mass motion of the quark system. In   
the s- and u-channel, ${\bf A}$ has the following expression for  
 different channels,    
\begin{eqnarray}    
s-channel:&&{\bf A}=-\frac{{\bf q}}{E_f+M_f},\\    
u-channel:&&{\bf A}=-(\frac{1}{E_f+M_f}+\frac{1}{E_i+M_i}){\bf k}-\frac{1}  
{E_f+M_f}{\bf q}.      
\end{eqnarray}    
    
The transverse transition amplitude for the s- and u-channel is,    
\begin{eqnarray}    
M^{s+u}_{fi}(T)&=&i\langle N_f|[g_e,H^T_m]|N_i\rangle \nonumber\\    
&&+i\omega\sum_{j}\langle N_f|H^T_m|N_j\rangle\langle N_j|\frac{1}{E_i+\omega-E_j}h_e|N_i\rangle\nonumber\\
&&+i\omega\sum_{j} \langle N_f|h_e\frac{1}{E_i-\omega_m-E_j}|N_j  
\rangle\langle N_j|H^T_m|N_i\rangle    
\label{4.3}    
\end{eqnarray}    
The nonrelativistic expansion of the first term gives,
\begin{eqnarray}    
M^{seagull}_{fi}(T)&=&-ig^s_3ae_m\{g_v\langle N_f|\{{\bf A}\cdot  
\veps_v,{\bf R}\cdot\veps\}|N_i \rangle \nonumber\\    
&&+g_v\langle N_f|\sum_{l}\{\frac{1}{2m_l}{\bf p}^\prime_l  
\cdot\veps_v,{\bf r}^\prime_l\cdot\veps\} |N_i \rangle\nonumber\\    
&&-\langle N_f|\{\frac{1}{E+M}+\frac{1}{2m_q}\}
\vsig\cdot(\veps_v\times\veps)|N_i \rangle\}  
e^{-\frac{({\bf k}-{\bf q})^2}{6\alpha^2}}    
\end{eqnarray}    
where $\{A,B\}=AB+BA$ is the anti-commutation operator. ${\bf R}$ is the coordinate of the center mass motion of the three quark system and ${\bf q}^\prime_l$, ${\bf r}^\prime_l $ denote the internal momentum and coordinate of the $l$th quark.

The seagull terms in the transverse  
 transitions are proportional to the charge of the outgoing mesons and, therefore,  
 vanish in neutral vector meson photoproduction. 

\subsection*{\bf 3.2. U-channel transition amplitudes}    

The last term in Eq.(\ref{4.2}) is the longitudinal transition amplitude  
 in the u-channel. We find
\begin{eqnarray}    
M^u_{fi}(L)&=&i\omega\sum_{j} \langle N_f|h_e\frac{1}{E_i-\omega_m-E_j}|N_j\rangle\langle N_j  
|-\frac{\mu}{|{\bf q}|}J_0|N_i\rangle\nonumber\\    
&=& (M^u_3+M^u_2)e^{-\frac{{\bf q}^2+{\bf k}^2}{6\alpha^2}}    
\end{eqnarray}    
in the harmonic oscillator basis, where    
\begin{eqnarray}    
M^u_3&=&g^u_3 \frac{a\mu}{|{\bf q}|}\{\frac{i}{2m_q}\vsig\cdot(\veps\times{\bf k})F^0  
(\frac{{\bf k}\cdot {\bf q}}{3\alpha^2},P_f\cdot k)\nonumber\\    
&&-g_v\frac{\omega}{3\alpha^2}{\bf q}\cdot\veps F^1  
(\frac{{\bf k}\cdot {\bf q}}{3\alpha^2},P_f\cdot k)\},    
\label{5.2}    
\end{eqnarray}    
which corresponds to incoming photons and outgoing vector mesons 
being absorbed and emitted by the same quark, and    
\begin{eqnarray}    
M^u_2&=&g^u_2 \frac{a\mu}{|{\bf q}|}\{g^\prime_a\frac{i}{2m_q}\vsig\cdot(\veps\times{\bf k})  F^0(-\frac{{\bf k}\cdot {\bf q}}{6\alpha^2},P_f\cdot k)\nonumber\\    
&&+g^\prime_v\frac{\omega}{6\alpha^2}{\bf q}\cdot\veps   
F^1(-\frac{{\bf k}\cdot {\bf q}}{6\alpha^2},P_f\cdot k)\}    
\label{5.3}
\end{eqnarray}    
in which the incoming photons and outgoing vector mesons are absorbed and
emitted by different quarks. $P_f$ in Eq.(\ref{5.2}) and Eq.(\ref{5.3}) 
denotes the four momentum of the final state nucleon.
 The function $F$ in Eq.(\ref{5.2}) and Eq.(\ref{5.3}) is defined as,    
\begin{equation}    
F^l(x,y)=\sum_{n\ge l}\frac{M_n}{(n-l)!(y+n\delta M^2)}x^{n-l},    
\end{equation}    
where $n\delta M^2=(M^2_n-M^2_f)/2$ represents 
the average mass difference between the ground state and excited states 
with the total excitation quantum number $n$ in the harmonic 
oscillator basis. The parameter $\alpha^2$ in the above equation is commonly 
used in the quark model and is related to the harmonic oscillator strength.   
    
Similarly, the transverse transition in the u-channel is given by,
\begin{eqnarray}    
M^u_{fi}(T)&=&i\omega\sum_{j} \langle N_f|h_e\frac{1}{E_i-\omega_m-E_j}|N_j\rangle\langle N_j|H^T_m|N_i\rangle\nonumber\\    
&=& (M^u_3+M^u_2)e^{-\frac{{\bf q}^2+{\bf k}^2}{6\alpha^2}}    
\end{eqnarray}    
where     
\begin{eqnarray}    
M^u_3/g^u_3&=&\frac{b^\prime}{4m^2_q}\{g_v(\veps\times
{\bf k})\cdot({\bf   
q}\times\veps_v) +i\vsig\cdot(\veps\times{\bf k})\times({\bf q}  
\times\veps_v) \}F^0(\frac{{\bf k}\cdot {\bf q}}{3\alpha^2},P_f\cdot k)  
\nonumber\\    
&&-\frac{ia}{2m_q}\vsig\cdot(\veps\times{\bf k}){\bf A}\cdot\veps_v   
F^0(\frac{{\bf k}\cdot {\bf q}}{3\alpha^2},P_f\cdot k)\nonumber\\    
&&+\{\frac{ia}{12m^2_q}\vsig\cdot(\veps\times{\bf k})\veps_v\cdot{\bf   
k}+\frac{ib^\prime\omega}{6m_q\alpha^2}\vsig\cdot({\bf q}\times\veps_v)  
\veps\cdot{\bf q}\nonumber\\    
&&+g_v\frac{a\omega}{3\alpha^2}\veps \cdot{\bf q}{\bf A}\cdot\veps_v-   
g_v\frac{a\omega}{6m_q}\veps \cdot\veps_v \} F^1(\frac{{\bf k\cdot   
q}}{3\alpha^2},P_f\cdot k)\nonumber\\    
&&-g_v\frac{a\omega}{18m_q\alpha^2}\veps_v\cdot{\bf k}\veps  
\cdot{\bf q} F^2(\frac{{\bf k}\cdot {\bf q}}{3\alpha^2},P_f\cdot k)    
\label{amp3}    
\end{eqnarray}    
and    
\begin{eqnarray}    
M^u_2/g^u_2&= &\frac{b^\prime}{4m^2_q}\{g^\prime_v(\veps\times{\bf k})  
\cdot({\bf  
 q}\times\veps_v)\nonumber\\
&& +ig^\prime_a\vsig\cdot(\veps\times{\bf k})
\times({\bf q}\times\veps_v) \}F^0(-\frac{{\bf k}\cdot {\bf q}}  
{6\alpha^2},P_f\cdot k)  
\nonumber\\    
&&-\frac{ia}{2m_q}\vsig\cdot(\veps\times{\bf k}){\bf A}\cdot\veps_v   
F^0(-\frac{{\bf k}\cdot {\bf q}}{6\alpha^2},P_f\cdot k)\nonumber\\    
&&+\{-\frac{ia}{24m^2_q}\vsig\cdot(\veps\times{\bf k})\veps_v\cdot{\bf   
k}-\frac{ib^\prime\omega}{12m_q\alpha^2}\vsig\cdot({\bf q}\times\veps_v)  
\veps\cdot{\bf q}\nonumber\\    
&&-g^\prime_v\frac{a\omega}{6\alpha^2}\veps \cdot{\bf q}{\bf A}  
\cdot\veps_v- g^\prime_v\frac{a\omega}{12m_q}\veps \cdot\veps_v \}   
F^1(-\frac{{\bf k}\cdot {\bf q}}{6\alpha^2},P_f\cdot k)\nonumber\\    
&&-g^\prime_v\frac{a\omega}{72m_q\alpha^2}\veps_v\cdot{\bf k}  
\veps\cdot{\bf q} F^2(-\frac{{\bf k}\cdot {\bf q}}{6\alpha^2},P_f\cdot k)    
\label{amp2}    
\end{eqnarray}    
The $g$-factors in Eq.(\ref{5.2})-(\ref{amp2})
are defined as     
\begin{equation}     
\langle N_f|\sum_{j} {\hat I}_j \vsig_j|N_i\rangle=g_A \langle N_f|   
\vsig|N_i\rangle,    
\end{equation}      
\begin{equation} g^u_3=\langle N_f|\sum_{j} e_j {\hat
I}_j\sigma^z_j|N_i\rangle/g_A,    
\end{equation}    
\begin{equation} g^u_2=\langle N_f|\sum_{i\neq j} e_j {\hat   
I}_i\sigma^z_j|N_i\rangle/g_A,    
\end{equation}    
\begin{equation} g_v=\langle N_f|\sum_{j} e_j{\hat I}_j  
|N_i\rangle/g^u_3g_A,    
\end{equation}       
\begin{equation}    
g^\prime_v=\frac{1}{3g^s_2g_A}\langle N_f|\sum_{i\neq j}  
 {\hat I}_ie_j\vsig_i\cdot\vsig_j|N_i\rangle,    
\end{equation}    
\begin{equation}    
g^\prime_a=\frac{1}{2g^s_2g_A}\langle N_f|\sum_{i\neq j}   
{\hat I}_ie_j(\vsig_i\times\vsig_j)_z|N_i\rangle.    
\end{equation}    
The numerical values of these $g$-factors have been derived in 
Ref.~\cite{pseudoscalar} in the $SU(6)\otimes O(3)$ symmetry limit;
they are listed in Table 1 for completeness. 

The first terms of Eq.(\ref{amp3}) and Eq.(\ref{amp2}) correspond  
 to the correlation between the magnetic transition and the c.m.
motion of the meson transition operator and they contribute to the  
 leading Born term in the u-channel. The second terms are due to
 correlations between the internal and c.m. motion of the photon   
and meson transition operators, and they only contribute to the  
 ground and $n\ge 1$ excited states in the harmonic oscillator  
 basis. The last terms in both equations represent the correlation
 of the internal motion between the photon and meson transition   
operators, which only contribute to transitions between the  
 ground and $n\ge 2$ excited states. 

As pointed out before, the mass splitting between 
the ground state spin 1/2 and spin 3/2  is significant, the transition
amplitudes for $\Delta$ resonance in $\rho$ production or $\Sigma^*$
resonance in $K^*$ production have to be computed separately.  
The transition amplitude with $n=0$ corresponding to the correlation
of magnitic transitions is,    
\begin{eqnarray}    
M^u(n=0)&=&-\frac{1}{2m_q}\frac{Me^{-\frac{{\bf q}^2+{\bf k}^2}
{6\alpha^2}}}{P_f\cdot k+\delta M^2/2}    
\frac{b^\prime}{2m_q}\{(g^u_3g_v+g^u_2g^\prime_v)({\bf k}\times\veps)  
\cdot({\bf q}\times\veps_v)\nonumber\\    
&&-i(g^u_3+g^u_2g^\prime_a)\vsig\cdot({\bf k}\times\veps)
\times({\bf q}\times\veps_v)\}.    
\end{eqnarray}    
The amplitude for spin 1/2 intermediate states in the total $n=0$  
 amplitudes is,    
\begin{eqnarray}    
&  &\langle N_f|h_e |N(J=1/2)\rangle\langle N(J=1/2)|H_m|N_i \rangle  
\nonumber\\    
&=  &\frac{\mu_N b^\prime}{2m_q}\frac{M_fe^{-\frac{{\bf q}^2+{\bf k}^2}{6\alpha^2}}}  
{P_f\cdot k+\delta M^2/2}    
\{ ({\bf k}\times\veps)\cdot({\bf q}\times\veps_v)\nonumber\\    
&&+i\vsig\cdot({\bf k}\times\veps)\times({\bf q}\times\veps_v)\}    
\end{eqnarray}    
 where $\mu_N$ is the magnetic moment, which has the following   
values for different processes,    
\begin{equation}    
\mu_N =\left\{ \begin{array}{ccc}    
\mu_\Lambda+\frac{g_{K^*\Sigma N}}{g_{K^*\Lambda N}}\mu_{\Lambda\Sigma}   
  &\qquad\mbox{for}\qquad  & \gamma N  
\to K^*\Lambda\\    
\mu_{\Sigma^0}+\frac{g_{K^*\Lambda N}}{g_{K^*\Sigma N}}  
\mu_{\Lambda\Sigma}    &\qquad\mbox{for}\qquad &  
 \gamma N\to K^*\Sigma \\
\mu_{N_f}& \qquad\mbox{for}\qquad    &\gamma N  
\to \rho N_f     
\end{array} \right.    
\end{equation}    

Thus, we obtain the spin 3/2 resonance contribution to the transition   
amplitude by subtracting the spin 1/2 intermediate state contributions  
 from the total $n=0$ amplitudes as follows:
\begin{eqnarray}    
M^u&= &-\frac{b^\prime}{2m_q}\frac{M_fe^{-\frac{{\bf q}^2+{\bf k}^2}{6\alpha^2}}}  
{P_f\cdot k+\delta M^2/2}    
\{[(g^u_3g_v+g^u_2g^\prime_v)/2m_q+\mu_N]({\bf k}\times\veps)  
\cdot({\bf q}\times\veps_v)\nonumber\\    
&&+i[(g^u_3+g^u_2g^\prime_a)/2m_q+\mu_N]\vsig\cdot[({\bf   
k}\times\veps)\times({\bf q}\times\veps_v)]\}.
\end{eqnarray}    
Substituting the $g$-factor coefficients into the above equation 
gives the following general expression for spin 3/2 resonance with $n=0$,
\begin{eqnarray}    
M^u&= &-\frac{b^\prime}{2m_q}\frac{M_fg_se^{-\frac{{\bf q}^2
+{\bf k}^2}{6\alpha^2}}}  
{M_N(P_f\cdot k+\delta M^2/2)}
\{2\vsig\cdot({\bf q}\times\veps_v)\vsig\cdot({\bf k}\times\veps)\nonumber\\
&&+i\vsig\cdot[({\bf q}\times\veps_v)\times({\bf k}\times\veps)]\}    
\end{eqnarray}
where the value of $g_s$ is given in Table 1.

Note that the transition amplitudes here are generally written as 
operators that are similar to the CGLN amplitudes in pseudoscalar 
meson photoproduction. They have to be transformed into the helicity 
amplitudes defined in Eq.(\ref{helicity}). In Tables 2 and 3,
 we show the relations between the operators presented here and 
the helicity amplitudes; they
are generally related by the Wigner $d$-function. 
  
\subsection*{\bf 3.3 S-channel transition amplitudes }    
    
The third term in Eq.(\ref{4.2}) and second term in Eq.(\ref{4.3})  
 are the s-channel longitudinal and transverse transition amplitudes.   
 Following the derivation for Compton scattering  
 in Ref.~\cite{Compton}, we obtain the general transition amplitude for 
excited states in the s-channel,    
\begin{equation}  
H^J_{a\lambda_V}=\frac{2M_R}
{s-M_R(M_R-i\Gamma({\bf q}))}   
h^J_{a\lambda_V},    
\label{6.1}    
\end{equation}    
where $\sqrt{s}=E_i+\omega=E_f+\omega_m$ is the total energy   
of the system, and $H^J_{a\lambda_V}$ are the helicity amplitudes
defined above. $\Gamma({\bf q})$ in Eq. \ref{6.1} denotes the  
 total width of the resonance,   which is
a function of the final state momentum ${\bf q}$.  For a resonance   
decaying into a two-body final state with relative angular momentum $l$,  
the decay width $\Gamma({\bf q})$ is  given by:
\begin{equation}\label{40}  
\Gamma({\bf q})= \Gamma_R \frac {\sqrt {s}}{M_R} \sum_{i} x_i   
\left (\frac {|{\bf q}_i|}{|{\bf q}^R_i|}\right )^{2l+1}   
\frac {D_l({\bf q}_i)}{D_l({\bf q}^R_i)},  
\end{equation}  
with   
\begin{equation}\label{41}  
|{\bf q}^R_i|=\sqrt{\frac   
{(M_R^2-M_N^2+M_i^2)^2}{4M_R^2}-M_i^2},  
\end{equation}  
and
\begin{equation}\label{42}  
|{\bf q}_i|=\sqrt{\frac   
{(s-M_N^2+M_i^2)^2}{4s}-M_i^2},  
\end{equation}  
where $x_i$ is the branching ratio of the resonance decaying into a   
meson with mass $M_i$ and a nucleon, and $\Gamma_R$ is the total   
decay width   
of the resonance with the mass $M_R$.  The   
function $D_l({\bf q})$ in Eq. \ref{40}, called fission barrier~\cite{bw}, 
is wavefunction dependent and has the following form
in the harmonic oscillator basis: 
\begin{equation}\label{43}  
D_l({\bf q})=exp\left (-\frac {{\bf q}^2}{3\alpha^2}\right ),  
\end{equation}  
which is independent of $l$. In principle, the branching ratio  
$x_i$ should also be evaluated in the quark model.    

For a given intermediate resonance state with spin $J$, the four   
independent helicity amplitudes $h^J_{a\lambda_V}$ in  Eq.(\ref{6.1})  
are a combination of the meson and photon helicity amplitudes  
 together with the Wigner-$d$ functions  
\begin{equation}
h^J_{a\lambda_V}=\sum_{\Lambda_f}d^J_{\Lambda_f,\Lambda_i}(\theta)  
A^V_{\Lambda_f}A^\gamma_{\Lambda_i},    
\label{59}
\end{equation}    
where $\Lambda_f=\lambda_V-\lambda_2$, $\Lambda_i=\lambda-\lambda_1$ 
and ${\bf k}\cdot{\bf q}=|{\bf k}||{\bf q}|cos(\theta)$.

The $A^\gamma_{1/2}$ and $A^\gamma_{3/2}$ in Eq.(\ref{59})  
represent the helicity amplitudes in the s-channel for the photon  
interactions; their explicit expressions have been given 
in Ref.~\cite{thesis}.

More explicitly, the 12 independent helicity amplitudes are related to
 the photon helicity amplitudes $A^\gamma_{\frac 12}$, $A^\gamma_{\frac 32}$ 
and vector meson helicity amplitudes $S^V_{\frac 12}$, 
$A^V_{\frac 12}$ and $A^V_{\frac 32}$ through the following relations

\begin{eqnarray}    
h^J_{11}&=&d^J_{\frac{1}{2},\frac{3}{2}}(\theta)A^V_{\frac{1}{2}}  
A^\gamma_{\frac{3}{2}},\nonumber\\    
h^J_{10}&=&d^J_{-\frac{1}{2},\frac{3}{2}}(\theta)S^V_{-\frac{1}{2}}
A^\gamma_{\frac{3}{2}},\nonumber\\    
h^J_{1-1}&=&d^J_{-\frac{3}{2},\frac{3}{2}}(\theta)A^V_{-\frac{3}{2}}    
A^\gamma_{\frac{3}{2}}
\end{eqnarray}    
for $a=1$, and $\lambda_V=1,0,-1$,
\begin{eqnarray}    
h^J_{21}&=&d^J_{\frac{1}{2},\frac{1}{2}}(\theta)A^V_{\frac{1}{2}}  
A^\gamma_{\frac{1}{2}},\nonumber\\    
h^J_{20}&=&d^J_{-\frac{1}{2},\frac{1}{2}}(\theta)S^V_{-\frac{1}{2}}    
A^\gamma_{\frac{1}{2}},\nonumber\\    
h^J_{2-1}&=&d^J_{-\frac{3}{2},\frac{1}{2}}(\theta)A^V_{-\frac{3}{2}}    
A^\gamma_{\frac{1}{2}}   
\end{eqnarray}    
for $a=2$, and $\lambda_V=1,0,-1$, 
\begin{eqnarray}    
h^J_{31}&=&d^J_{\frac{3}{2},\frac{3}{2}}(\theta)A^V_{\frac{3}{2}}    
A^\gamma_{\frac{3}{2}},\nonumber\\    
h^J_{30}&=&d^J_{\frac{1}{2},\frac{3}{2}}(\theta)S^V_{\frac{1}{2}}    
A^\gamma_{\frac{3}{2}},\nonumber\\    
h^J_{3-1}&=&d^J_{-\frac{1}{2},\frac{3}{2}}(\theta)A^V_{-\frac{1}{2}}    
A^\gamma_{\frac{3}{2}}  
\end{eqnarray}
for $a=3$, and $\lambda_V=1,0,-1$,
and 
\begin{eqnarray}    
h^J_{41}&=&d^J_{\frac{3}{2},\frac{1}{2}}(\theta)A^V_{\frac{3}{2}}   
A^\gamma_{\frac{1}{2}},\nonumber\\   
h^J_{40}&=&d^J_{\frac{1}{2},\frac{1}{2}}(\theta)S^V_{\frac{1}{2}}    
A^\gamma_{\frac{1}{2}},\nonumber\\    
h^J_{4-1}&=&d^J_{-\frac{1}{2},\frac{1}{2}}(\theta)A^V_{-\frac{1}{2}}    
A^\gamma_{\frac{1}{2}}   
\end{eqnarray}    
for $a=4$, and $\lambda_V=1,0,-1$.

The amplitudes with negative helicities in the above equations
 are not independent   
from those with positive one;  they are related by 
an additional phase factor  according to the Wigner-Eckart  
 theorem,     
\begin{equation}    
A^V_{-\lambda}= (-1)^{J_f-J_i-J_V}A^V_{\lambda}    
\end{equation}    
where $J_f$ and $J_i$ are the final nucleon and initial  
 resonance spins, and $J_V$ is the angular momentum of the vector meson.
The angular distributions of the helicity amplitudes in terms of the multipole
transitions have been discussed in Ref. \cite{yang}, the expressions
 here are consistent  with their analysis.

The evaluation of the vector meson helicity amplitudes are 
similar to that of the photon amplitudes. The transition operator for
a resonance decaying into a vector meson and a nucleon is,
\begin{equation}    
H^T_m=\sum_{l} \{i\frac{b^\prime}{2m_q}\vsig_l\cdot({\bf q}
\times\veps_v) +\frac{a}{2\mu_q}{\bf p}^\prime_l  
\cdot\veps_v\}{\hat I}_le^{-i{\bf q\cdot r}_l},     
\end{equation}  
for transverse transitions and
\begin{equation}
H^L_m=\frac{a\mu}{|{\bf q}|}\sum_{l} {\hat I}_le^{-i{\bf q\cdot r}_l}
\end{equation}
for longitudinal transitions. Thus, $H^T_m$ and $H^L_m$ have 
the group structure,
\begin{equation}
H^T_m={\hat I}_3(A L^-_{(3)} +B \sigma^-_{(3)}),
\label{htm}
\end{equation}
and 
\begin{equation}
H^L_m={\hat I}_3 S,
\label{hlm}
\end{equation}
where
\begin{equation}
\label{A}
A=\frac{3 a}{2\sqrt{2}m_q}\langle\psi_f|p^-_3
e^{-i{\bf q}\cdot{\bf r}_3}|\psi_R\rangle,
\end{equation}
\begin{equation}
B=\frac{-3 b^\prime}{2m_q}
|{\bf q}|\langle\psi_f|e^{-i{\bf q}\cdot{\bf r}_3}|\psi_R\rangle,
\end{equation}
\begin{equation}
S=-\frac{3\mu a}{|{\bf q}|}\langle\psi_f|
e^{-i{\bf q}\cdot{\bf r}_3}|\psi_R\rangle.
\end{equation}
where $p^-_3=p_x-ip_y$. 
In Eq.(\ref{htm}), $L^-_{(3)}$ and $\sigma^-_{(3)}$ denote orbital and spin flip operators.
The helicity amplitudes $A^V_{\frac 12}$, $A^V_{\frac 32}$
and $S^V_{\frac 12}$ are the matrix elements of Eq.(\ref{htm}) and Eq.(\ref{hlm}).
We list the angular momentum and flavor parts of $A^V_{\frac 12}$, $A^V_{\frac 32}$ 
and $S^V_{\frac 12}$ for $\omega$ and $\rho$ photoproduction
 in Tables 4-6 in the 
$SU(6)\otimes O(3)$ limit with $A$, $B$ and $S$ in the second row to denote the corresponding  spatial integrals, which are given in Table 7.

The resonances with $n\ge 3$ are treated as degenerate since there is
little information available about them. Their longitudinal
 transition in the s-channel is given by:
\begin{equation}
h^J_{a\lambda_V=0}=(M^s_3(L)+M^s_2(L))
e^{-\frac{{\bf q}^2+{\bf k}^2}
{6\alpha^2}}
\end{equation}
where 
\begin{eqnarray}
M^s_3(L)&=&g^s_3 \frac{a\mu}{|{\bf q}|}\{-\frac{i}{2m_q}\vsig\cdot(\veps\times{\bf k}) 
\frac{1}{n!}(\frac{{\bf k}\cdot {\bf q}}{3\alpha^2})^n\nonumber\\
&& +g_v\frac{\omega}{3\alpha^2}{\bf q}\cdot\veps\frac{1}{(n-1)!}
(\frac{{\bf k\cdot 
q}}{3\alpha^2})^{n-1}\},
\label{s-l-3}
\end{eqnarray}
and
\begin{eqnarray}
M^s_2(L)&=&-g^u_2 \frac{a\mu}{|{\bf q}|}
\{g^\prime_a\frac{i}{2m_q}\vsig\cdot(\veps\times{\bf k})
\frac{1}{n!}(\frac{-{\bf k\cdot 
q}}{6\alpha^2})^n\nonumber\\
&&+g^\prime_v\frac{\omega}{6\alpha^2}{\bf q}\cdot\veps\frac{1}{(n-1)!}(\frac{-{\bf k\cdot 
q}}{6\alpha^2})^{n-1}\}.
\label{s-l-2}
\end{eqnarray}
The $g$-factors in Eq.(\ref{s-l-3}) and (\ref{s-l-2})
have been defined previously, and 
\begin{equation}
g^s_3=\langle N_f|\sum_{j} {\hat I}_j e_j\sigma^z_j|N_i\rangle/g_A=
e_m+g^u_3,
\end{equation}
where $e_m$ is the charge of the outgoing vector meson.

The transverse transition amplitudes at the quark level are:
\begin{equation}
h^J_{a\lambda_V=\pm 1}=(M^s_3(T)+M^s_2(T))
e^{-\frac{{\bf q}^2+{\bf k}^2}{6\alpha^2}}
\end{equation}
where 
\begin{eqnarray}
M^s_3(T)/g^s_3&=&\frac{b^\prime}{4m^2_q}\{g_v({\bf q}\times\veps_v)\cdot(\veps\times{\bf k})
+i\vsig\cdot({\bf q}\times\veps_v)\times(\veps\times{\bf k})\}\frac{1}{n!}(\frac{{\bf k\cdot 
q}}{3\alpha^2})^n\nonumber\\
&&+\{-\frac{ia}{12m^2_q}\vsig\cdot(\veps\times{\bf k})\veps_v
\cdot{\bf k}
+\frac{ib^\prime\omega}{6m_q\alpha^2}\vsig\cdot({\bf q}\times\veps_v)\veps\cdot{\bf q}\nonumber\\
&&+g_v\frac{a\omega}{6m_q}\veps_v\cdot\veps\}\frac{1}{(n-1)!}
(\frac{{\bf k}\cdot {\bf q}}{3\alpha^2})^{n-1}\nonumber\\
&&+g_v\frac{a\omega}{18m_q\alpha^2}\veps_v\cdot{\bf k}\veps
\cdot{\bf q} \frac{1}{(n-2)!}
(\frac{{\bf k}\cdot{\bf q}}{3\alpha^2})^{n-2},
\end{eqnarray}
and
\begin{eqnarray}
M^s_2(T)/g^u_2&=&\frac{b^\prime}{4m^2_q}\{g^\prime_v({\bf q}\times\veps_v)\cdot(\veps\times{\bf 
k})+ig^\prime_a\vsig\cdot({\bf q}\times\veps_v)\times(\veps\times{\bf k})\}\frac{1}{n!}(\frac{-{\bf 
k\cdot q}}{6\alpha^2})^n\nonumber\\
&&+\{\frac{ia}{24m^2_q}\vsig\cdot(\veps\times{\bf k})\veps_v\cdot{\bf
k}-\frac{ib^\prime\omega}{12m_q\alpha^2}\vsig\cdot({\bf q}\times\veps_v)\veps\cdot{\bf q}\nonumber\\
&&-g^\prime_v\frac{a\omega}{12m_q}\veps_v\cdot\veps \}\frac{1}{(n-1)!}(\frac{-{\bf k\cdot 
q}}{6\alpha^2})^{n-1}\nonumber\\
&&+g^\prime_v\frac{a\omega}{72m_q\alpha^2}\veps_v\cdot{\bf k}\veps
\cdot{\bf q} \frac{1}{(n-2)!}(\frac{-{\bf k}\cdot {\bf q}}{6\alpha^2})^{n-2}.
\end{eqnarray}

Qualitatively, we find that the resonances with larger partial waves 
have larger decay widths into the vector meson and nucleon though this is 
not as explicit as in the pseudoscalar case~\cite{pseudoscalar,eta}. Thus, we could use the mass
and decay width of the high spin states, such as $G_{17}(2190)$ for $n=3$ 
states and $H_{19}(2220)$ for $n=4$ states in the $\omega$ photoproduction. 
The relation between these operators and the helicity amplitudes
 $h_{a\lambda_V}$ has been given in Table 2 and 3.

\section*{\bf 4. Discussion and conclusion}    

In this paper we have developed the framework and formalism
for the description on vector meson photoproduction in the 
constituent quark model.
The use of an effective Lagrangian allows gauge invariance to be
satisfied straightforwardly.
The advantage of using the quark model approach is that
the number of free parameters is greatly reduced in comparison
to hadronic models that introduce each resonance as a new
independent field with unknown coupling constants.
In our approach,
there are only three parameters in the $SU(6)\otimes O(3)$ symmetry limit, 
the coupling constants $a$ and $b$ which determine the
coupling strengths of the vector meson to the quark,
and the harmonic oscillator strength $\alpha^2$. 

One significant approximation inherent in the presented approach
is the treatment of the vector mesons as point particles, thus,
the effects due to the finite size 
of the vector mesons that were important in the $^3P_0$ model
are neglected here.
A possible way that may partly compensate this problem 
is adjusting the parameter $\alpha^2$,  the harmonic
oscillator strength. In general, the question of how to include the finite 
size of vector mesons 
while maintaining gauge invariance is 
very complicated and has not yet been resolved.

While this approach should give a qualitative description
of vector meson photoproduction, a precise quantitative agreement with 
the data cannot be expected, as configuration mixing effects for the 
resonances in the second and third resonance region are known to be very important. 
Such effects could be investigated in our approach by inserting a 
mixing parameter $C_R$ in front of the transition amplitudes for the  s-channel 
resonances, as was done in Ref.~\cite{eta}. 
Yet even without new parameters the present approach should be able
to answer basic questions such as the need for an additional pomeron
or other t-channel exchanges in light of the duality argument.
We believe that the model presented here provides 
a systematic method to investigate 
the resonance behavior in the vector meson production for the first time.
It therefore can assist in answering lingering quetions about the
existence of the ``missing resonances``.
The numerical implementation of this approach for
$\rho$ and $\omega$ photoproduction is in progress and will be published elsewhere.

The authors acknowledge useful discussions with B. Saghai, F. Tabakin, P. Cole
and F.J. Klein.
This work is supported in part by Chinese Education Commission and Peking 
University, and the US-DOE grant no. DE-FG02-95-ER40907.

\section*{\bf Appendix}    

The matrix element for the nucleon pole term of transverse 
excitations in the s-channel is,
\begin{eqnarray}
M^s_N(T)&=&-\frac{M_Ne^{-\frac{{\bf q}^2+{\bf k}^2}{6\alpha^2}}}{P_N\cdot k}\{g^t_v\frac{\omega ae_N}{E_f+M_f}\veps_v\cdot\veps -g_A\mu_N\frac{b^\prime}{2m_q}[(\veps_v\times {\bf q})\cdot(\veps
\times {\bf k})\nonumber\\
&&+i\vsig\cdot(\veps_v\times {\bf q})\times(\veps\times {\bf k})]\},
\end{eqnarray}
while the one for the u-channel is,
\begin{eqnarray}
M^u_N(T) & = & -\frac{M_fe^{-\frac{{\bf q}^2
+{\bf k}^2}{6\alpha^2}}}{P_f\cdot k}\{ g^t_v\frac{\omega ae_f}{E_N+M_N}\veps\cdot\veps_v
+g_A\mu_f\frac{b^\prime}{2m_q}
\big [(\veps\times {\bf k})\cdot(\veps_v\times {\bf q})
  \nonumber  \\ 
& &  + i\vsig\cdot((\veps\times {\bf k})\times(\veps_v\times {\bf q}))
\big ]\}\nonumber\\
&&+ \frac{ e_f e^{-\frac{{\bf q}^2+{\bf k}^2}{6\alpha^2}}}
{P_f\cdot k}\{\frac{-g^t_v a}{E_N+M_N}{\bf q}\cdot\veps{\bf k}\cdot\veps_v
 +ig_A\frac{b^\prime}{2m_q}\vsig\cdot(\veps_v\times {\bf q})
{\bf q}\cdot\veps\}. 
\end{eqnarray}

The matrix element for the nucleon pole term of the longitudinal 
excitations in the s-channel is,
\begin{equation}
M^s_N(L)=-g^t_v\frac{i\mu a }{|{\bf q}|}
\frac{M_N}{P_N\cdot k}\mu_N\vsig\cdot(\veps\times{\bf k})e^{-\frac{{\bf q}^2
+{\bf k}^2}{6\alpha^2}},
\end{equation}
while the one for the u-channel is, 
\begin{equation}
M^u_N(L)=g^t_v\frac{\mu a }{|{\bf q}|}\frac{1}{P_f\cdot k}\{ -e_f{\bf q}
\cdot \veps+iM_f\mu_f\vsig\cdot(\veps\times{\bf k})\}
e^{-\frac{{\bf q}^2+{\bf k}^2}{6\alpha^2}},
\end{equation}
where the $g$-factor $g^t_v$ has the following form,
\begin{equation}
g^t_v=\langle N_f|\sum_{j}{\hat I}_j|N_i\rangle.
\end{equation}

The t-channel matrix element for the transverse transition is,
\begin{eqnarray}
M^t(T)&=&\frac{a e_m}{q\cdot k}\{-g^t_v[\omega_m+(\frac{{\bf q}}{E_f+M_f}+\frac{{\bf k}}{E_N+M_N})\cdot{\bf q}]\veps\cdot\veps_v
\nonumber\\
&&+g_A\frac{i}{2m_q}\vsig\cdot({\bf k}\times{\bf q})\veps\cdot\veps_v\nonumber\\
&&-g^t_v(\frac{1}{E_f+M_f}+\frac{1}{E_N+M_N}){\bf q}\cdot\veps{\bf k}\cdot\veps_v \nonumber\\
&&+g_A\frac{i}{2m_q}\vsig\cdot(({\bf k}-{\bf q})\times\veps_v){\bf q}\cdot\veps\nonumber\\
&&+g_A\frac{i}{2m_q}\vsig\cdot(({\bf k}-{\bf q})\times\veps){\bf k}\cdot\veps_v
 \}e^{-\frac{({\bf k}-{\bf q})^2}{6\alpha^2}},
\end{eqnarray}
and for the longitudinal transition is,
\begin{equation}
M^t(L)=\frac{\mu}{|{\bf q}|}\frac{ae_m}{q\cdot k}\{g^t_v(1-\frac{\omega}{E_f+M_f}){\bf q}\cdot\veps
+g_A\frac{i\omega}{2m_q}\vsig\cdot(({\bf k}-{\bf q})\times\veps) \}e^{-\frac{({\bf k}-{\bf q})^2}{6\alpha^2}}.
\end{equation}

\vfill

\newpage

\begin{tabular}{lccccccccc}
\multicolumn{10}{l}
{ Table 1: The $g$-factors in the u-channel amplitudes in Eqs. }\\
\multicolumn{10}{l}
{\ref{amp3} and \ref{amp2} for different production processes.}\\[1ex]
\hline\\
\hline
Reactions & & $g^u_3$ & $g^u_2$ & $g_v$ & $g_v^\prime$& $g_a^\prime$ 
& $g_A$ & $g_S$&$g^t_v$\\[1ex]\hline
$\gamma p\to K^{*+}\Lambda$ & & -$\frac 13$ &  $\frac 13$ & 1 & 1 
&1& $\sqrt{\frac 32}$ & -$\frac {\mu_{\Lambda}}3$&$\sqrt{\frac 32}$\\[1ex]
$\gamma n\to K^{*0}\Lambda$ & & -$\frac 13$ & $\frac 13$ & 1 & -1 
&-1&  $\sqrt{\frac 32}$ & $\frac {\mu_{\Lambda}} 3$&$\sqrt{\frac 32}$\\[1ex]
$\gamma p \to K^{*+}\Sigma^0$ & & -$\frac 13$ & $\frac 13$ & -3 & -7
& 9 & -$\frac 1{3\sqrt{2}}$ & $\mu_{\Sigma^0}$&$\frac{1}{\sqrt 2}$\\[1ex]
$\gamma n\to K^{*0}\Sigma^0$ & & -$\frac 13$ & $\frac 13$ & -3 & 11 
&-9& $\frac 1{3\sqrt{2}}$ &  $\mu_{\Sigma^0}$&$-\frac{1}{\sqrt 2}$\\[1ex]
$\gamma p\to K^{*0}\Sigma^+$ & & -$\frac 13$ & $\frac 43$ & -3 & 2 
&0 & $\frac 13$ &  $\frac {2\mu_{\Sigma^+}}{3}$ &0\\[1ex]
$\gamma n\to K^{*+}\Sigma^-$ & & -$\frac 13$& -$\frac 23$ & -3 & 2 
&0&  -$\frac 13$ &  0&1\\[1ex]
$\gamma p\to \omega p$ & & 1 & 0 & 1 & 0 & 0&  1 & 0 &3\\[1ex]
$\gamma n\to \omega n$ & &  -$\frac 23$  & $\frac 23$ & 0 
& -1&  0&  1 & 0 &3\\[1ex]
$\gamma p\to \rho^+ n$ & & -$\frac 13$  & $\frac 13$ & $\frac 35$ 
&$\frac 15$ & $\frac 95$ & $\frac 53$ & -$\frac{2\mu_n}5$ &1\\[1ex]
$\gamma n\to \rho^- p$ & & $\frac 23$ & $\frac 13$ & $\frac 35$ 
&$\frac 15$ &- $\frac 95$  &  - $\frac 53$ & -$\frac{4\mu_p}{15}$&1\\[1ex]
$\gamma p\to \rho^0 p$ & & $\frac 7{15}$ & $\frac 8{15}$ 
&$\frac {15}7$ &  2 & 0 &  $\frac 5{3\sqrt{2}}$ & $\frac{8\mu_p}{15}$
&$-\frac{1}{\sqrt 2}$\\[1ex]
$\gamma n\to \rho^0 n$ & & -$\frac 2{15}$ & $\frac 2{15}$ & 6 &  -7 
& 0 & -$\frac 5{3\sqrt{2}}$ & $\frac{4\mu_n}5$&$\frac{1}{\sqrt 2}$
\\[1ex]\hline
\end{tabular}    

\vspace{1.0cm}
\begin{tabular}{lll}
\multicolumn{3}{l}
{ Table 2: The operators in the longitudinal excitations expressed }\\
\multicolumn{3}{l}
{in terms of the helicity amplitudes. ${\hat{\bf k}}$ and ${\hat{\bf q}}$ are the unit vectors}\\
\multicolumn{3}{l}
{of ${\bf k}$ and ${\bf q}$, respectively. Other components of $H_{a\lambda_V}$ are 
zero. }\\
\multicolumn{3}{l}
{ The $d$ functions have the rotation angle $\theta$. }\\ [1ex]
\hline\hline
Operators&$H_{10}(\lambda_2=1/2) $&$H_{20}(\lambda_2=1/2) $\\
&$H_{30}(\lambda_2=-1/2) $&$H_{40}(\lambda_2=-1/2) $
\\[1ex]\hline
${\hat{\bf q}}\cdot \veps$
&$d^1_{10}d^{\frac 12}_{-\frac 12\lambda_2 }$
&$d^1_{10} d^{\frac 12}_{\frac 12\lambda_2 } $
\\[1ex]\hline
$\vsig\cdot(\veps\times{\hat{\bf k}})$
&$i\sqrt{2} d^{\frac 12}_{\frac 12\lambda_2 }$
&0\\[1ex]\hline
$\vsig\cdot(\veps\times{\hat{\bf q}})$
&$ id^1_{10} d^{\frac 12}_{-\frac 12\lambda_2 }$
&$-id^1_{10} d^{\frac 12}_{\frac 12\lambda_2 }$\\
&$+i\sqrt{2}d^1_{00}d^{\frac 12}_{\frac 12\lambda_2 } $& 
\\[1ex]\hline\hline
\end{tabular}    

\begin{tabular}{lll}
\multicolumn{3}{l}
{ Table 3: The operators in the transverse excitations expressed in terms of
the }\\
\multicolumn{3}{l}
{helicity amplitudes. ${\hat{\bf k}}$ and ${\hat{\bf q}}$ are the unit vectors of ${\bf k}$ and ${\bf 
q}$, respectively. }\\
\multicolumn{3}{l}
{ The $d$ functions have the rotation angle $\theta$. 
$\lambda_V=\pm 1$ denote the helicities  }\\
\multicolumn{3}{l}
{of the mesons, and the vector ${\hat{\bf Z}}$ is defined as ${\hat{\bf Z}}=(\veps\times{\hat{\bf 
k}})
\times(\veps_v\times{\hat{\bf q}})$. }\\[1ex]
\hline\hline    
Operators
&$H_{1\lambda_V}(\lambda_2=1/2)$
&$H_{2\lambda_V}(\lambda_2=1/2)$\\
&$H_{3\lambda_V}(\lambda_2=-1/2)$
&$H_{4\lambda_V}(\lambda_2=-1/2)$\\[1ex] \hline

$(\veps\times{\hat{\bf k}})\cdot(\veps_v\times{\hat{\bf q}})$
&$-\lambda_V d^1_{1\lambda_V} d^{\frac 12}_{-\frac 12\lambda_2}$
&$-\lambda_V d^1_{1\lambda_V} d^{\frac 12}_{\frac 12\lambda_2 }$
\\[1ex]\hline

$\vsig\cdot{\hat{\bf Z}}$
&$-i\lambda_V d^1_{11} d^{\frac 12}_{-\frac 12\lambda_2 }$
&$i\lambda_V d^1_{11} d^{\frac 12}_{\frac 12\lambda_2 }$\\
&$-i\sqrt{2}\lambda_V d^1_{0\lambda_V} d^{\frac 12}_{\frac 12\lambda_2 }$
&\\[1ex]\hline

$ \vsig \cdot(\veps\times{\hat{\bf k}}){\hat{\bf k}}\cdot\veps_v $
&$i\sqrt{2}d^1_{0\lambda_V} d^{\frac 12}_{\frac 12\lambda_2 }$&0
\\[1ex]\hline

$ \vsig \cdot(\veps_v\times{\hat{\bf q}}){\hat{\bf q}}\cdot\veps $
&$i\sqrt{2} \lambda_Vd^1_{10}d^1_{1\lambda_V} d^{\frac 12}_{\frac 12\lambda_2 }$
&$i\sqrt{2} \lambda_Vd^1_{10}d^1_{-1\lambda_V} d^{\frac 12}_{-\frac 12\lambda_2 }$\\
&$-i \lambda_Vd^1_{10}d^1_{0\lambda_V} d^{\frac 12}_{-\frac 12\lambda_2 }$
&$+i \lambda_Vd^1_{10}d^1_{0\lambda_V} d^{\frac 12}_{\frac 12\lambda_2 }$
\\[1ex]\hline

$\veps_v\cdot\veps$
&$ d^1_{1\lambda_V} d^{\frac 12}_{-\frac 12\lambda_2 }$
&$d^1_{1\lambda_V} d^{\frac 12}_{\frac 12\lambda_2 }$\\[1ex]\hline

${\hat{\bf q}}\cdot\veps {\hat{\bf k}}\cdot\veps_v $
&$d^1_{0\lambda_V}d^1_{01} d^{\frac 12}_{-\frac 12\lambda_2 }$
&$d^1_{0\lambda_V}d^1_{01} d^{\frac 12}_{\frac 12\lambda_2 }$
\\[1ex]\hline

$ \vsig \cdot(\veps\times{\hat{\bf q}}){\hat{\bf k}}\cdot\veps_v $
&$i\sqrt{2}d^1_{00}d^1_{0\lambda_V} d^{\frac 12}_{\frac 12\lambda_2 }$&\\
&$+id^1_{10}d^1_{0\lambda_V} d^{\frac 12}_{\frac 12\lambda_2 }$
&$-id^1_{10}d^1_{0\lambda_V} d^{\frac 12}_{\frac 12\lambda_2 }$
\\[1ex]\hline

$ \vsig \cdot(\veps_v\times{\hat{\bf k}}){\hat{\bf q}}\cdot\veps $
&$-i\sqrt 2d^1_{10}d^1_{-1\lambda_V} d^{\frac 12}_{\frac 12\lambda_2 }$
&$-i\sqrt 2d^1_{10}d^1_{1\lambda_V} d^{\frac 12}_{-\frac 12\lambda_2 }$
\\[1ex]\hline

$\vsig\cdot(\veps_v\times\veps)$
&$-i\sqrt 2d^1_{0\lambda_V} d^{\frac 12}_{\frac 12\lambda_2 }$&\\
&$-id^1_{1\lambda_V} d^{\frac 12}_{-\frac 12\lambda_2 }$
&$id^1_{1\lambda_V} d^{\frac 12}_{\frac 12\lambda_2 }$
\\[1ex]\hline

$\vsig \cdot({\hat{\bf k}}\times{\hat{\bf q}})\veps\cdot\veps_v $
&$i\sqrt 2d^1_{-10}d^1_{1\lambda_V} d^{\frac 12}_{\frac 12\lambda_2 }$
&$i\sqrt 2d^1_{10}d^1_{1\lambda_V} d^{\frac 12}_{-\frac 12\lambda_2 }$
\\[1ex]\hline\hline    
\end{tabular}

\vfill
\newpage
    
\begin{tabular}{lcccccccccc}
\multicolumn{11}{l}
{ Table 4:  The agular momentum and flavor parts of the helicity amplitudes  }\\
\multicolumn{11}{l}
{ for $\gamma p\to \omega p$, or $\gamma n\to \omega n$, in the $SU(6)\otimes O(3)$ symmetry limit. They are }\\
\multicolumn{11}{l}
{ the coefficients of the spatial integrals $A$, $B$ and $S$ in Eq.(\ref{htm}) and (\ref{hlm}). }\\
\multicolumn{11}{l}
{ The analytic expressions for $A$, $B$ and $S$ in Table 4-6 are given in Table 7. }\\
\multicolumn{11}{l}
{ The ``*`` in Table 4-6 denotes those states decoupling in spin and flavor space,}\\
\multicolumn{11}{l}
{ thus, their amplitudes are zero.}\\[1ex]    
\hline\hline    
States & & $S_{1/2}$ & & &$A_{1/2}$ & & & & $A_{3/2}$  &\\[1ex] \hline     
 & & S & & A & & B & & A & & B\\[1ex] \hline    
$N(^2P_M)_{\frac 12^-}$&&0&&0&&$-\frac 2{3\sqrt{6}}$&&0&&0\\[1ex]    
$ N(^2P_M)_{\frac 32^-}$&&0&&0&&$\frac 2{3\sqrt{3}}$&&0&&0\\[1ex]    
$ N(^4P_M)_{\frac 12^-}$&&0&&0&&$-\frac 1{3\sqrt{6}}$&&*&&*\\[1ex]    
$ N(^4P_M)_{\frac 32^-}$&&0&&0&&$\frac 1{3\sqrt{30}}$&&0&&$-\frac1{\sqrt{15}}$\\[1ex]    
$ N(^4P_M)_{\frac 52^-}$&&0&&0&&$\frac 1{\sqrt{30}}$&&0&&$\frac1{\sqrt{10}}$\\[1ex]    
$N(^2D_S)_{\frac 32^+}$&&$-\sqrt{\frac 25}$&&$\sqrt{\frac 35}
$&&$-\frac 13 \sqrt{\frac 25}$&&$-\frac1{\sqrt{5}}$&&*\\[1ex]    
$N(^2D_S)_{\frac 52^+}$&&$\sqrt{\frac 35}$&&$\sqrt{\frac 25}
$&&$\frac 13 \sqrt{\frac 35}$&&$\frac2{\sqrt{5}}$&&*\\[1ex]    
$N(^2S^\prime_S)_{\frac 12^+}$&&1&&*&&$\frac 13$&&*&&*\\[1ex]    
$N(^2S_M)_{\frac 12^+}$&&0&&*&&$\frac 2{3\sqrt{2}}$&&*&&*\\[1ex]    
$N(^4S_M)_{\frac 32^+}$&&0&&*&&$\frac 1{3\sqrt{2}}$&&*&&$\frac 1{\sqrt{6}}$\\[1ex]
$N(^2D_M)_{\frac 32^+}$&&0&&0&&$-\frac 2{3\sqrt{5}}$&&0&&*\\[1ex]    
$N(^2D_M)_{\frac 52^+}$&&0&&0&&$\sqrt{\frac 2{15}}$&&0&&*\\[1ex]    
$N(^4D_M)_{\frac 12^+}$&&0&&0&&$\frac 1{3\sqrt{10}}$&&*&&*\\[1ex]    
$N(^4D_M)_{\frac 32^+}$&&0&&0&&$-\frac 1{3\sqrt{10}}$&&0&&$\frac1{\sqrt{30}}$\\[1ex]    
$N(^4D_M)_{\frac 52^+}$&&0&&0&&$-\frac 1{\sqrt{210}}$&&0&&
$-\sqrt{\frac 3{35}}$\\[1ex]
$N(^4D_M)_{\frac 72^+}$&&0&&0&&$\frac 1{\sqrt{35}}$&&0&&$\frac 1{\sqrt{21}}$\\[1ex]
\hline    
\end{tabular}

\begin{tabular}{lcccccccccc}
\multicolumn{11}{l}
{ Table 5:  The angular momentum and flavor parts of the helicity amplitudes }\\
\multicolumn{11}{l}
{ for $\gamma p\to \rho^0 p$ in the $SU(6)\otimes O(3)$ symmetry limit, while those }\\
\multicolumn{11}{l}
{ for $\gamma n\to\rho^0 n$ are given by $A(\gamma n\to \rho^0 n)=
(-1)^{I+1/2}A(\gamma p\to \rho^0 p)$, where }\\
\multicolumn{11}{l}
{ $I$ is the isospin of the resonances.}\\ [1ex]
\hline\hline    
States & & $S_{1/2}$ & & &$A_{1/2}$ & & & & $A_{3/2}$  
&\\[1ex] \hline     
 & & S & & A & & B & & A & & B\\[1ex]\hline    
$N(^2P_M)_{\frac 12^-}$&&$\frac 1{3\sqrt{3}}$&&$-\frac{\sqrt 2}{3\sqrt3}$&&$\frac 2{9\sqrt{3}}$&&*&&*\\[1ex]    
$ N(^2P_M)_{\frac 32^-}$&&$-\frac 13\sqrt{\frac 23}$&&$-\frac 1{3\sqrt{3}}$&&$-\frac 29\sqrt{\frac 23}$&&$-\frac 13$&&*\\[1ex]    
$ N(^4P_M)_{\frac 12^-}$&&0&&0&&$-\frac 1{18\sqrt{3}}$&&*&&*\\[1ex]    
$ N(^4P_M)_{\frac 32^-}$&&0&&0&&$\frac 1{18\sqrt{15}}$    
&&0&&$\frac 1{6\sqrt5}$\\[1ex]    
$ N(^4P_M)_{\frac 52^-}$&&0&&0&&$\frac 1{6\sqrt{15}}$&&0&&$
\frac 1{\sqrt{30}}$\\[1ex]    
$\Delta(^2P_M)_{\frac 12^-}$&&$-\frac 1{3\sqrt3}$&&$\frac 13
\sqrt{\frac 23}$&&$\frac 1{9\sqrt3}$&&*&&*\\[1ex]    
$\Delta(^2P_M)_{\frac 32^-}$&&$\frac 13\sqrt{\frac 23}$&&$\frac 1{3\sqrt3}$&&$-\frac 19\sqrt{\frac 23}$&&$\frac 13$&&*\\[1ex]    
$N(^2D_S)_{\frac 32^+}$&&$\frac 1{3\sqrt5}$&&$-\frac 1{\sqrt{30}}$
&&$\frac 1{9\sqrt{5}}$&&$\frac1{3\sqrt{10}}$&&*\\[1ex]    
$N(^2D_S)_{\frac 52^+}$&&$-\frac 1{\sqrt{30}}$&&$-\frac 1{3\sqrt5}$
&&$-\frac 5{3\sqrt{30}}$&&$-\frac2{3\sqrt{10}}$&&*\\[1ex]    
$\Delta(^4D_S)_{\frac 12^+}$&&0&&0&&$\frac 2{9\sqrt5}$&&*&&*\\[1ex]    
$\Delta(^4D_S)_{\frac 32^+}$&&0&&0&&$-\frac 2{9\sqrt5}$&&0&&$\frac 2{3\sqrt{15}}$\\[1ex]    
$\Delta(^4D_S)_{\frac 52^+}$&&0&&0&&$-\frac 29\sqrt{\frac 3{35}}$
&&0&&$-\frac 23\sqrt{\frac 6{35}}$\\[1ex]    
$\Delta(^4D_S)_{\frac 72^+}$&&0&&0&&$\frac 23\sqrt{\frac2{35}}$
&&0&&$\frac 4{3\sqrt{42}}$\\[1ex]    
$N(^2S^\prime_S)_{\frac 12^+}$&&$-\frac1{3\sqrt2}$&&*&&$-\frac 5{9\sqrt2}$&&*&&*\\[1ex]    
$\Delta(^4S^\prime_S)_{\frac 32^+}$&&0&&*&&$\frac 29$&&*&&$\frac 2{3\sqrt3}$\\[1ex]    
$\Delta(^4S_S)_{\frac 32^+}$&&0&&*&&$\frac 29$&&*&&
$\frac 2{3\sqrt3}$\\[1ex]    
$N(^2S_M)_{\frac 12^+}$&&$-\frac 13$&&*&&$-\frac 29$&&*&&*\\[1ex]    
$N(^4S_M)_{\frac 32^+}$&&0&&*&&$\frac 1{18}$&&*&&
$\frac 1{6\sqrt{3}}$\\[1ex]    
$\Delta(^2S_M)_{\frac 12^+}$&&$\frac 13$&&*&&$-\frac 19
$&&*&&*\\[1ex]
$N(^2D_M)_{\frac 32^+}$&&$\frac 13\sqrt{\frac 25}$&&
$-\frac 13\sqrt{\frac 35}$&&$\frac 29\sqrt{\frac 25}$&&
$\frac 1{3\sqrt 5}$&&*\\[1ex]    
$N(^2D_M)_{\frac 52^+}$&&$-\frac 13\sqrt{\frac 35}$&&
$-\frac 13\sqrt{\frac 25}$&&$-\frac 29\sqrt{\frac 35}$&&
$-\frac 2{3\sqrt 5}$&&*\\[1ex]    
$N(^4D_M)_{\frac 12^+}$&&0&&0&&$\frac 1{18\sqrt{5}}$&&*&&*\\[1ex]    
$N(^4D_M)_{\frac 32^+}$&&0&&0&&$-\frac 1{18\sqrt{5}}$&&0&&$\frac1{6\sqrt{15}}$\\[1ex]    
$N(^4D_M)_{\frac 52^+}$&&0&&0&&$-\frac 1{18}\sqrt{\frac 3{35}}$
&&0&&$-\frac 16\sqrt{\frac 6{35}}$\\[1ex]
$N(^4D_M)_{\frac 72^+}$&&0&&0&&$\frac 1{3\sqrt{70}}$&&0&&$\frac 1{3\sqrt{42}}$\\[1ex]
$\Delta(^2D_M)_{\frac 32^+}$&&$-\frac 13\sqrt{\frac 25}$&&
$\frac 13\sqrt{\frac 35}$&&$\frac 19\sqrt{\frac 25}$&&
$-\frac 1{3\sqrt 5}$&&*\\[1ex]
$\Delta(^2D_M)_{\frac 52^+}$&&$\frac 13\sqrt{\frac 35}$&&
$\frac 13\sqrt{\frac 25}$&&$-\frac 19\sqrt{\frac 35}$&&
$\frac 2{3\sqrt 5}$&&*\\[1ex]\hline    
\end{tabular}

\begin{tabular}{lcccccccccc}
\multicolumn{11}{l}
{ Table 6:  The angular momentum and flavor parts of the helicity amplitudes }\\
\multicolumn{11}{l}
{ for $\gamma p\to \rho^+ n$ in the $SU(6)\otimes O(3)$ symmetry limit, while those  }\\
\multicolumn{11}{l}
{ for $\gamma n\to\rho^- p$ are given by $A(\gamma n\to \rho^- p)=(-1)^{I+1/2}
A(\gamma p\to \rho^+ n)$, where }\\
\multicolumn{11}{l}
{$I$ is the isospin of the resonances. }\\[1ex]      
\hline\hline    
States & & $S_{1/2}$ & & &$A_{1/2}$ & & & & $A_{3/2}$  
&\\[1ex] \hline     
 & & S & & A & & B & & A & & B\\[1ex]\hline    
$N(^2P_M)_{\frac 12^-}$&&$-\frac 2{3\sqrt{6}}$&&$\frac 
2{3\sqrt3}$&&$-\frac 4{9\sqrt{6}}$&&*&&*\\[1ex]    
$ N(^2P_M)_{\frac 32^-}$&&$\frac 2{3\sqrt{3}}$&&$
\frac 2{3\sqrt{6}}$&&$\frac 4{9\sqrt{3}}$&&$
\frac 2{3\sqrt 2}$&&*\\[1ex]    
$ N(^4P_M)_{\frac 12^-}$&&0&&0&&$\frac 1{9\sqrt{6}}
$&&*&&*\\[1ex]    
$ N(^4P_M)_{\frac 32^-}$&&0&&0&&$-\frac 1{9\sqrt{30}}$    
&&0&&$-\frac 1{3\sqrt{10}}$\\[1ex]    
$ N(^4P_M)_{\frac 52^-}$&&0&&0&&$-\frac 1{3\sqrt{30}}
$&&0&&$-\frac 1{3\sqrt{15}}$\\[1ex]    
$\Delta(^2P_M)_{\frac 12^-}$&&$-\frac 1{3\sqrt6}$&&
$\frac 1{3\sqrt3}$&&$\frac 1{9\sqrt6}$&&*&&*\\[1ex]    
$\Delta(^2P_M)_{\frac 32^-}$&&$\frac 1{3\sqrt3}$&&
$\frac 1{3\sqrt6}$&&$-\frac 1{9\sqrt3}$&&$\frac 1{3\sqrt2}$&&*\\[1ex]    
$N(^2D_S)_{\frac 32^+}$&&$-\frac 13\sqrt{\frac 25}$&&
$\frac 13\sqrt{\frac 35}$&&$-\frac 59\sqrt{\frac 25}$&&$-\frac1{3\sqrt{5}}$&&*\\[1ex]
$N(^2D_S)_{\frac 52^+}$&&$\frac 13\sqrt{\frac 35}$&&
$\frac 13\sqrt{\frac 25}$&&$\frac 59\sqrt{\frac 35}$&&$\frac2{3\sqrt{5}}$&&*\\[1ex]    
$\Delta(^4D_S)_{\frac 12^+}$&&0&&0&&$\frac 2{9\sqrt{10}}
$&&*&&*\\[1ex]
$\Delta(^4D_S)_{\frac 32^+}$&&0&&0&&$-\frac 2{9\sqrt{10}}
$&&0&&$\frac 13\sqrt{\frac 2{15}}$\\[1ex]
$\Delta(^4D_S)_{\frac 52^+}$&&0&&0&&$-\frac 19
\sqrt{\frac 6{35}}
$&&0&&$\frac 23\sqrt{\frac 3{35}}$\\[1ex]    
$\Delta(^4D_S)_{\frac 72^+}$&&0&&0&&$\frac 2{3\sqrt{35}}$&&0&&
$\frac 2{3\sqrt{21}}$\\[1ex]    
$N(^2S^\prime_S)_{\frac 12^+}$&&$\frac 13$&&*&&$\frac 59$&&*&&*\\[1ex]    
$\Delta(^4S^\prime_S)_{\frac 32^+}$&&0&&*&&$\frac 2{9\sqrt2}$&&*&&
$\frac 13\sqrt{\frac 23}$\\[1ex]
$\Delta(^4S_S)_{\frac 32^+}$&&0&&*&&$\frac 2{9\sqrt2}$&&*&&$
\frac 13\sqrt{\frac 23}$\\[1ex]    
$N(^2S_M)_{\frac 12^+}$&&$\frac 2{3\sqrt2}$&&*&&$\frac 4{9\sqrt2}$
&&*&&*\\[1ex]    $N(^4S_M)_{\frac 32^+}$&&0&&*&&$-\frac 1{9\sqrt2}$
&&*&&$-\frac 1{3\sqrt{6}}$\\[1ex]
$\Delta(^2S_M)_{\frac 12^+}$&&$\frac 1{3\sqrt2}$&&*
&&$-\frac 1{9\sqrt2}$&&*&&*\\[1ex]
$N(^2D_M)_{\frac 32^+}$&&$-\frac 23\sqrt{\frac 25}$&&$\sqrt{\frac 2{15}}$&&$-\frac 4{9\sqrt5}$&&$-\frac 13\sqrt{\frac 25}$&&*\\[1ex]    
$N(^2D_M)_{\frac 52^+}$&&$\sqrt{\frac 2{15}}$&&$\frac 2{3\sqrt5}$
&&$\frac 49\sqrt{\frac 3{10}}$&&$\frac 23\sqrt{\frac 25}$&&*\\[1ex]    
$N(^4D_M)_{\frac 12^+}$&&0&&0&&$-\frac 1{9\sqrt{10}}$&&*&&*\\[1ex]    
$N(^4D_M)_{\frac 32^+}$&&0&&0&&$\frac 1{9\sqrt{10}}$&&0&&$-\frac1{3\sqrt{30}}$\\[1ex]    
$N(^4D_M)_{\frac 52^+}$&&0&&0&&$\frac 1{9}\sqrt{\frac 3{70}}$&&0&&$
\frac 13\sqrt{\frac 3{35}}$\\[1ex]
$N(^4D_M)_{\frac 72^+}$&&0&&0&&$-\frac 1{3\sqrt{35}}$&&0&&$-\frac 1{3\sqrt{21}}$\\[1ex]
$\Delta(^2D_M)_{\frac 32^+}$&&$-\frac 1{3\sqrt5}$&&$
\frac 1{\sqrt{30}}$&&$\frac 1{9\sqrt5}$&&$-\frac 1{3\sqrt{10}}$&&*\\[1ex]
$\Delta(^2D_M)_{\frac 52^+}$&&$\frac 1{\sqrt{30}}$&&$
\frac 1{3\sqrt5}$&&$-\frac 19\sqrt{\frac 3{10}}$&&$\frac 13
\sqrt{\frac 25}$&&*\\[1ex]\hline    
\end{tabular}    
    
\begin{tabular}{lcl}    
\multicolumn{3}{l}    
{ Table 7:  The spatial integrals in the harmonic oscillator basis. } \\[1ex]    
\hline\hline 
Multiplet& & Expression\\[1ex] \hline   
$[70, 1^-]_1$ & &$A=\frac{3a}{2m_q\sqrt{3}}\alpha 
exp(-\frac{{\bf q}^2}{6\alpha^2})$ \\[1ex]
 & &$B=\frac{ b^\prime}{m_q}\sqrt{\frac 32}\frac{{\bf q}^2}{\alpha} 
exp(-\frac{{\bf q}^2}{6\alpha^2})$ \\[1ex]
 & &$S=\frac{\sqrt3 \mu a}{\alpha} 
exp(-\frac{{\bf q}^2}{6\alpha^2})$ \\[1ex]

$[56, 2^+]_2$ & &$A=-\frac{a}{2\sqrt{2}m_q}|{\bf q}| 
exp(-\frac{{\bf q}^2}{6\alpha^2})$ \\[1ex]
 & &$B=-\frac{ b^\prime}{2\sqrt{3}m_q}|{\bf q}|(\frac{\bf q}{\alpha})^2 
exp(-\frac{{\bf q}^2}{6\alpha^2})$ \\[1ex]
 & &$S=-\frac{ \mu a}{\sqrt6 |{\bf q}|}(\frac{\bf q}{\alpha})^2 
exp(-\frac{{\bf q}^2}{6\alpha^2})$ \\[1ex]

$[56, 0^+]_2$& &$B=\frac{ b^\prime}{2\sqrt 3m_q}|{\bf q}|
(\frac{\bf q}{\alpha})^2 exp(-\frac{{\bf q}^2}{6\alpha^2})$ \\[1ex]
 & &$S=\frac{ \mu a}{\sqrt6 |{\bf q}|}(\frac{\bf q}{\alpha})^2 
exp(-\frac{{\bf q}^2}{6\alpha^2})$ \\[1ex]

$[56, 0^+]_0$& &$B=\frac{3 b^\prime}{\sqrt2 m_q}|{\bf q}|
exp(-\frac{{\bf q}^2}{6\alpha^2})$ \\[1ex]
& &$S=\frac{3 \mu a}{|{\bf q}|}exp(-\frac{{\bf q}^2}{6\alpha^2})$ \\[1ex]

$[70, 0^+]_2$& &$B=-\frac{ b^\prime}{2\sqrt6 m_q}|{\bf q}|
(\frac{\bf q}{\alpha})^2 exp(-\frac{{\bf q}^2}{6\alpha^2})$ \\[1ex]
& &$S=-\frac{\mu a}{2\sqrt3|{\bf q}|}(\frac{\bf q}{\alpha})^2 
exp(-\frac{{\bf q}^2}{6\alpha^2})$ \\[1ex]

$[70, 2^+]_2$& &$A=\frac{a}{2\sqrt2 m_q}|{\bf q}|
exp(-\frac{{\bf q}^2}{6\alpha^2})$\\[1ex]
& &$B=\frac{ b^\prime}{2\sqrt{3}m_q}|{\bf q}|
(\frac{\bf q}{\alpha})^2 exp(-\frac{{\bf q}^2}{6\alpha^2})$ \\[1ex]
& &$S=\frac{\mu a}{\sqrt{6}|{\bf q}|} (\frac{\bf q}{\alpha})^2 
exp(-\frac{{\bf q}^2}{6\alpha^2})$ \\[1ex]
\hline    
\end{tabular}

\end{document}